\documentclass[reprint,cite,doublecol,figures,prb,amsmath,amssymb,superscriptaddress,showpacs]{revtex4-1}
\usepackage[dvips]{graphicx}
\DeclareGraphicsExtensions{.pdf} \graphicspath{{./figs/}}
\usepackage{color}
\usepackage{verbatim}
\usepackage[german,english]{babel}

\usepackage{dcolumn}
\usepackage{bm}

\newcommand{\mb}{\mathbf}
\newcommand{\mr}{\mathrm}

\newcommand{\KCSCl}{K$_2$CuSO$_4$Cl$_2$}
\newcommand{\KCSBr}{K$_2$CuSO$_4$Br$_2$}
\newcommand{\KCSClBr}{K$_2$CuSO$_4$Ha$_2$}
\newcommand{\KCSX}{K$_2$CuSO$_4$(Cl$_{1-x}$Br$_x$)$_2$}

\newcommand{\etal}{\textit{et al}.}
\newcommand{\K}{\,\mr{K}}
\newcommand{\mK}{\,\mr{mK}}
\newcommand{\T}{\,\mr{T}}

\newcommand{\meV}{\,\mr{meV}}
\newcommand{\Angstrom}{\,\mr{\AA}}
\newcommand{\kB}{k_\mr{B}}
\newcommand{\TN}{T_\mr{N}}
\newcommand{\muB}{\mu_\mr{B}}

\newcommand{\DM}{{Dzyaloshinskii-Moriya}}

\newcommand{\NSM}{Neutron Scattering and Magnetism, Laboratory for Solid State Physics, ETH Z\"urich, CH-8093 Z\"urich, Switzerland}

\newcommand{\LNS}{Laboratory for Neutron Scattering and Imaging, Paul Scherrer Insitut, CH-5232 Villigen, Switzerland}

\newcommand{\secref}[1]{Sec.\,\protect\ref{#1}}
\newcommand{\figref}[1]{Fig.\,\protect\ref{#1}}

\begin{document}


\title{Quantum spin chains with frustration due to \DM{} interactions}
\author{M.~H\"alg}
\thanks{Corresponding author}
\email{haelgma@phys.ethz.ch}
\affiliation{\NSM}
\author{W.~E.~A.~Lorenz}
\affiliation{\NSM}
\author{K.~Yu.~Povarov}
\affiliation{\NSM}
\affiliation{P. L. Kapitza Institute for Physical Problems, RAS, 119334 Moscow, Russia}
\author{M.~M\aa nsson}
\affiliation{\NSM}
\affiliation{\LNS}
\author{Y.~Skourski}
\affiliation{Hochfeld-Magnetlabor Dresden, Helmholtz-Zentrum Dresden-Rossendorf, D-01314 Dresden, Germany}
\author{A.~Zheludev}
\affiliation{\NSM}

\date{\today}

\begin{abstract}
The properties of two quantum spin chain materials, \KCSCl{} and \KCSBr{}, are studied by a variety of experimental techniques, including bulk measurements, neutron spectroscopy and ESR.
The hierarchy of relevant terms in the magnetic Hamiltonian is established.
It is shown that these two compounds feature substantial \DM{} (DM) interactions that are \emph{uniform} within each chain, but \emph{antiparallel} in adjacent chains.
The result is a peculiar type of frustration of inter-chain interactions, which leads to an unusual field-temperature phase diagram.

\end{abstract}

\maketitle

\section{\label{sec:Introduction}Introduction}

The $S=1/2$ Heisenberg chain is an outstanding and versatile model system in quantum many-body physics.
An increasing number of predominantly Cu$^{2+}$ compounds is known to exhibit magnetic properties that closely resemble the Heisenberg spin chain, hence giving access to the experimental study of spin chains with various perturbations.
Amongst these are frustrated intra-chain exchange interactions,\cite{Hase1993,Lorenz2009,Linarite2013} staggered $g$-tensors,\cite{Casola2012} staggered \DM{} (DM) interactions \cite{CuBenzoate1997} or disorder.\cite{Masuda2004,Shiroka2013,Simutis2013}
Isolated spin chains are quantum critical and do not show magnetic long range order.
However, any inter-chain coupling will drive the system away from the critical point and induce magnetic ordering at finite temperatures.\cite{Schulz1996}
While sufficiently strong inter-chain interactions may eventually suppress peculiar ground states of the isolated chain,\cite{Nishimoto2011} frustrated inter-chain couplings can also give rise to extraordinarily complex magnetic phase diagrams.\cite{Starykh2010}

Theoretical predictions for spin chains with uniform DM interaction include the restoration of spin-entanglement due to the reduction of collinear spin-alignement\cite{Kargarian2009} as well as detailed magnetic phase diagrams as function of anisotropy and field direction.\cite{Garate2010}
\textit{Uniform} DM coupling, which arises as a consequence of intra-chain frustration (inverse DM interaction) is known in a number of compounds and the emerging multiferroicity in these systems has triggered a large number of studies.\cite{Park2007,Naito2007,Seki2010}
However, only few spin chain compounds are known to date that display uniform DM interaction as a direct consequence of the crystallographic structure.\cite{Dzyaloshinski1958,Moriya1960}
These few systems also show rather complex Hamiltonians, with strong inter-chain coupling (Cs$_2$CuCl$_4$),\cite{Starykh2010} staggered $g$-tensors (KCuGaF$_6$),\cite{Umegaki2012} intra-chain frustration (TiOCl)\cite{Zakharov2006} or very small DM coupling (BaV$_3$O$_8$).\cite{Tsirlin2014}

In the present work a variety of experimental techniques is employed to investigate a family of spin chain materials with the general formula \KCSClBr{} where Ha~$=$~Cl or Br represents halogen atoms.
Their most intriguing property is the presence of a uniform DM component to the intra-chain exchange.
It is demonstrated that the DM interactions, though weak compared to intra-chain exchange, are considerable in comparison to inter-chain coupling in these materials.
The result is a unique type of geometric frustration that significantly impacts the $H$-$T$ phase diagram.

\section{Experimental details}

Single crystal samples of the title compounds are grown by slow evaporation or by the temperature-gradient method\cite{Yankova2012} from stoichiometric hydrous solutions of K$_2$SO$_4$ and the respective Cu$^{2+}$-halide.
For this study large, nicely facetted single crystals with a mass up to a few grams were synthesized.
Both compounds mainly grow as rhombi elongated along the $b$-axis with $(101)$ and $(10\bar{1})$ facets;
the Cl-compound as light-blue and the Br-compound as dark-green crystals.
The crystallographic structures have been examined on a \textit{Bruker} AXS single crystal X-ray diffractometer, analyzing more than 2000 reflections collected per sample.

All experimental studies reported in this review were performed on single crystal specimen of \KCSCl{} and \KCSBr{}.
Bulk properties of the compounds, i.e. magnetization and specific heat have been measured on commercial \textit{Quantum Design} MPMS and PPMS instruments in the temperature range of $1.8\K$ to $300\K$.
Low temperature specific heat data down to 50 mK have been obtained employing the \textit{Quantum Design} dilution insert for the PPMS.
Measurements of the actual addenda were performed right before sample mounting for all specific heat data taken.
The temperature rise per measured data point was $2\%$ of the sample temperature.

Magnetization data up to the saturation field have been taken for both compounds, employing the iHelium3 insert for the MPMS in case of \KCSCl{} and pulsed field for \KCSBr{}.
Pulsed field magnetization measurements were carried out at the Dresden High Magnetic Field Laboratory in a 50 T magnet.
The signal recorded with a compensated pick-up coil was integrated and normalized to magnetization data taken at the same temperature up to 14 T with a VSM insert for the PPMS.

Neutron scattering experiments have been performed on a 3.0~g single crystal of \KCSCl{} and two coaligned single crystals of \KCSBr{} of mass 3.4~g and 2.3~g.
The crystals were aligned using the ORION instrument at SINQ/PSI.
Inelastic neutron scattering data have been collected at the TASP three axes spectrometer at SINQ/PSI.
The spectrometer was operated fully focussed without collimation for \KCSBr{} and deploying only $80'$ collimation after the monochromator in case of \KCSCl{}.
The neutron final momentum  $k_f$ was fixed to $1.5\Angstrom^{-1}$ in both cases.
An LN$_2$ cooled Be-filter was mounted to reduce higher order contamination.
For the experiment on \KCSCl{} a vertical field cryomagnet was used whereas \KCSBr{} was measured in a standard Orange cryostat.

ESR spectra have been recorded versus temperature at cavity resonance frequency $\nu\approx 27$ GHz in a home-built rectangular-cavity spectrometer at the P.~L.~Kapitza Institute for physical problems RAS.

\section{Experimental results}

\subsection{Crystallographic structure}\label{sec::structure}
While magnetic properties of the title compounds \KCSCl{} and \KCSBr{} have hitherto not been investigated, the chlorine compound is known as a mineral already since 1872.\cite{Scacchi1872}
Its crystallographic structure has been solved by 1976.\cite{Giacovazzo1976}
The compound crystallizes in orthorhombic $Pnma$ ($Z=4$) structure.
From the present X-ray diffraction data lattice constants $(a, b, c)_\mr{Cl} = (7.73(1), 6.08(1), 16.29(1))\Angstrom$ were obtained.
The second title compound, \KCSBr{}, grows in the same structure with a slightly enlarged unit cell $(a, b, c)_\mr{Br} = (7.73(1), 6.30(1), 16.43(1))\Angstrom$.

\begin{figure}[htb!]
\unitlength1cm
\includegraphics[width=.4\textwidth]{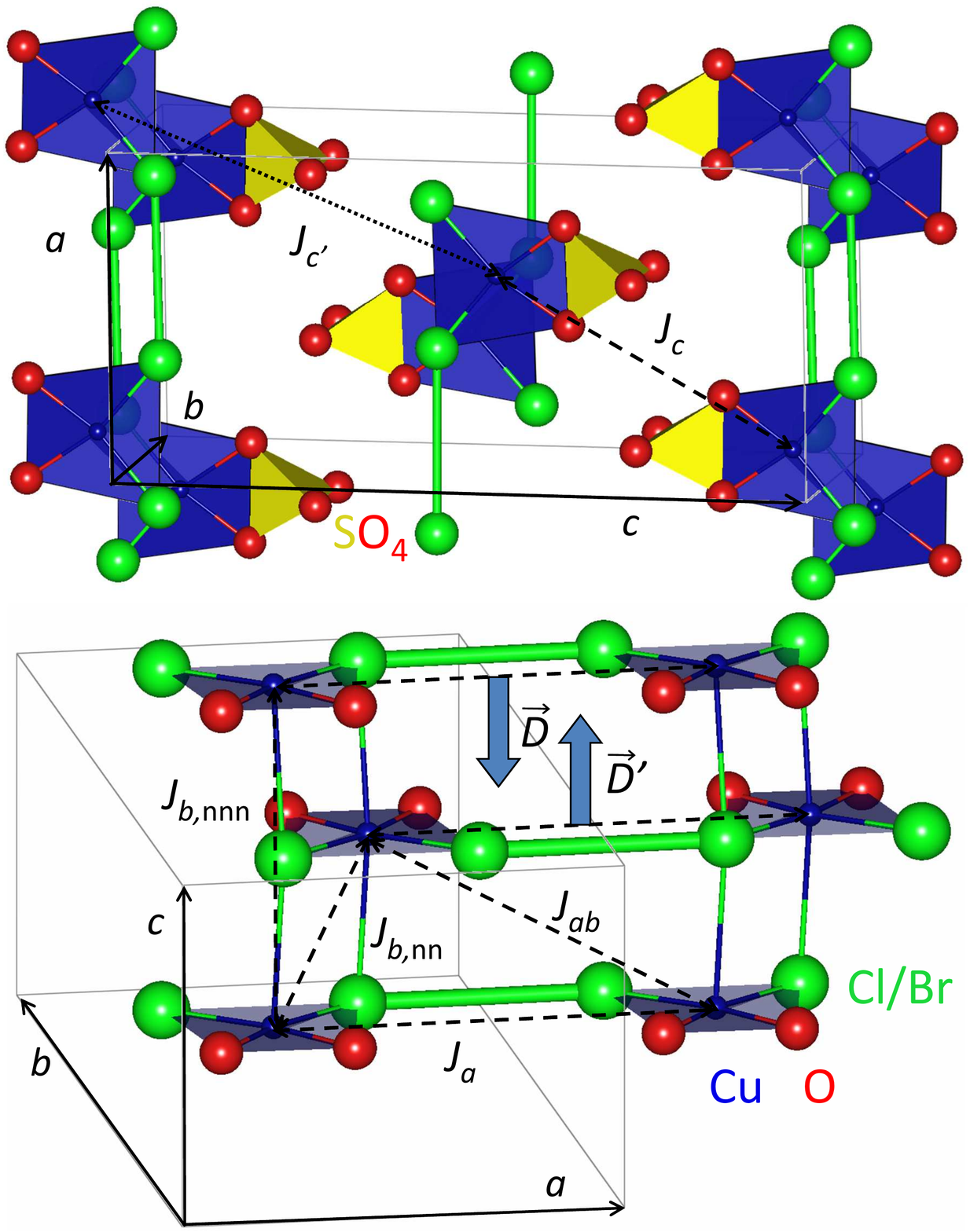}
\caption{Sketch of the crystallographic structure of the title compounds. K$^+$ ions are omitted for clarity. The dominant super-exchange path between two Cu$^{2+}$ ions, $J_a$, involves two halide ions, giving rise to spin chains along the $a$-axis. The contribution of the DM vector $\textbf{D}$ parallel to the $b$-axis is discussed in \secref{sec::DM}.  \label{fig::structure}}
\end{figure}

Details of the crystallographic structure are depicted in~\figref{fig::structure}. The materials are composed of CuSO$_4$Ha$_2^{2-}$ complexes separated along the $c$-axis by K$^{+}$-ions. CuO$_2$Ha$_2$ plaquettes are located in a mirror-plane perpendicular to the $b$-axis. The crystal structure suggests two potential predominant magnetic exchange patterns. Both the nearest and the second nearest neighbor Cu$^{2+}$-ions are located along the $b$-axis and might yield a frustrated spin chain along $b$. However, these exchange paths are mediated by apical bonds only and hence are likely weak. In contrast, the third nearest neighbor exchange path is mediated by equatorial Cu-Ha-Ha-Cu bonds and should give rise to spin chains reaching along the $a$-axis.
There are two distinct exchange paths along the $c$-axis that are both expected to be weak due to their long distance and the involvement of multiple hoppings.
For magnetic field applied along the crystallographic axes, all Cu$^{2+}$ show the same $g$-factor by symmetry.
However, for magnetic field with non-zero components along both the $a$- and $c$-axis, there will be a weakly alternating $g$-factor in neighboring chains along the $c$-axis.

Furthermore, it could be shown that the crystallographic structure of the title compounds is also stable for arbitrary site substitution on the halogen site, and compounds \KCSX{} can be grown for any $x\in [0,1]$.
Also, Na$_2$CuSeO$_4$Cl$_2$ proved to grow in the same structure.
Along with the strong dependence of the intra-chain coupling on chemical composition as shown below these compounds are highly suitable for the study of bond disorder in spin chain systems.

\subsection{Thermodynamic characterization}
\begin{figure}[htb!]
\unitlength1cm
\includegraphics[width=.48\textwidth]{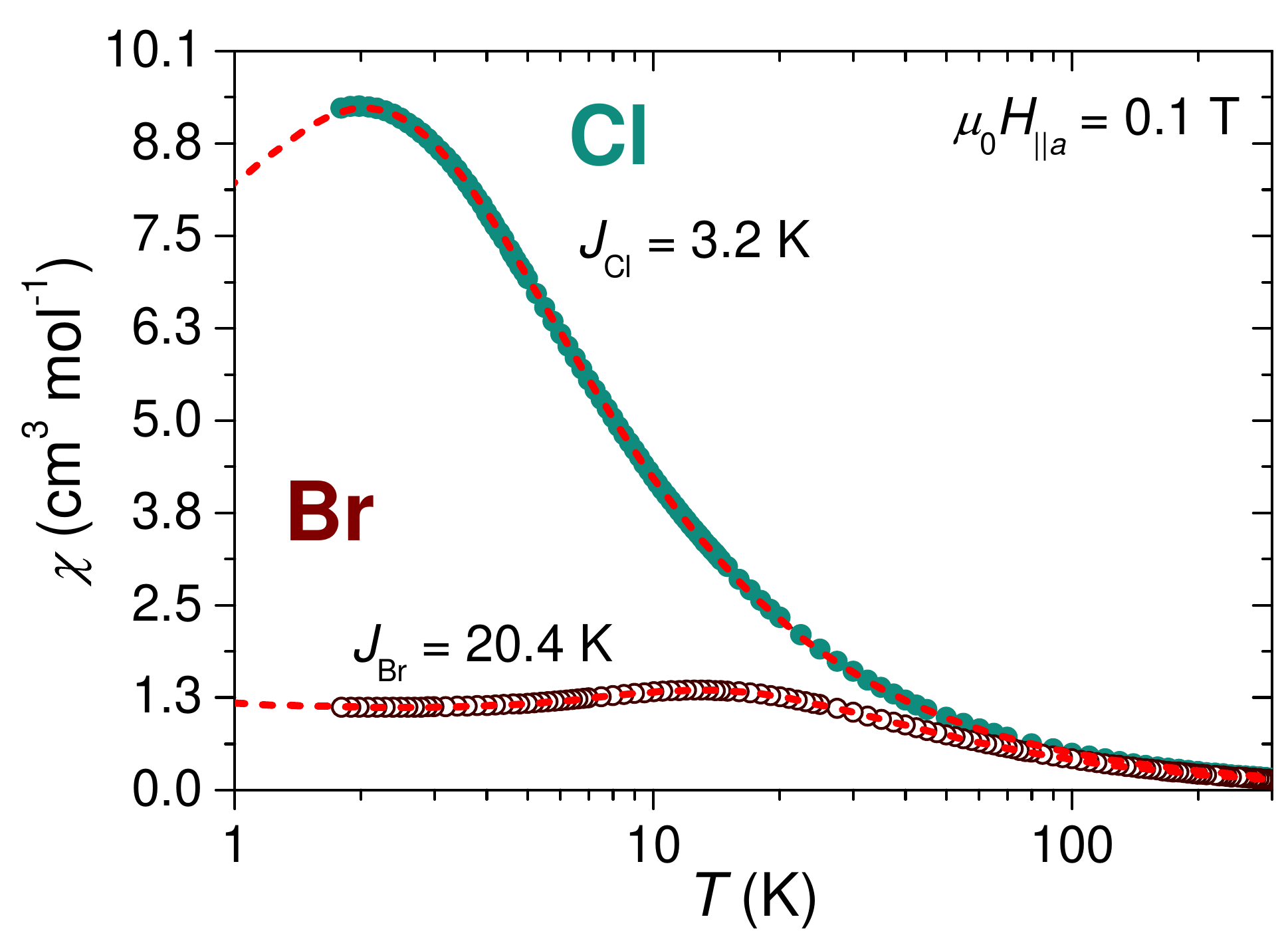}
\caption{The temperature dependence of DC-susceptibility measured in magnetic field of 0.1~T applied along the crystallographic $a$-axis is shown for both compounds. Dashed lines correspond to the spin chain susceptibility as described in the text.\label{fig::Mandcp-M}}
\end{figure}

Magnetic susceptibility as well as magnetic specific heat of both compounds exhibit broad maxima characteristic of low-dimensional systems.
These maxima are located at $1.99\K$ and $12.6\K$ in the magnetic susceptibility for the chlorine and bromine compound, respectively (\figref{fig::Mandcp-M}).
Using the empirical fitting function to the spin chain susceptibility by Johnston \etal{} \cite{Johnston2000} the data are well described with intra-chain exchange interactions $J_\text{Cl}=3.2(1)\K$ and $J_\text{Br}=20.4(1)\K$, respectively.
In the chlorine material the precision of the fit is restricted by the data reaching only down to $1.8\K$, i.e. just below the maximum. In the bromine material the fit is slightly improved assuming an additional Curie-like impurity contribution of $0.3\%$ of the total number of spins as may be attributed to surface contamination of this water-soluble compound.
\begin{figure}[htb!]
\unitlength1cm
\includegraphics[width=.48\textwidth]{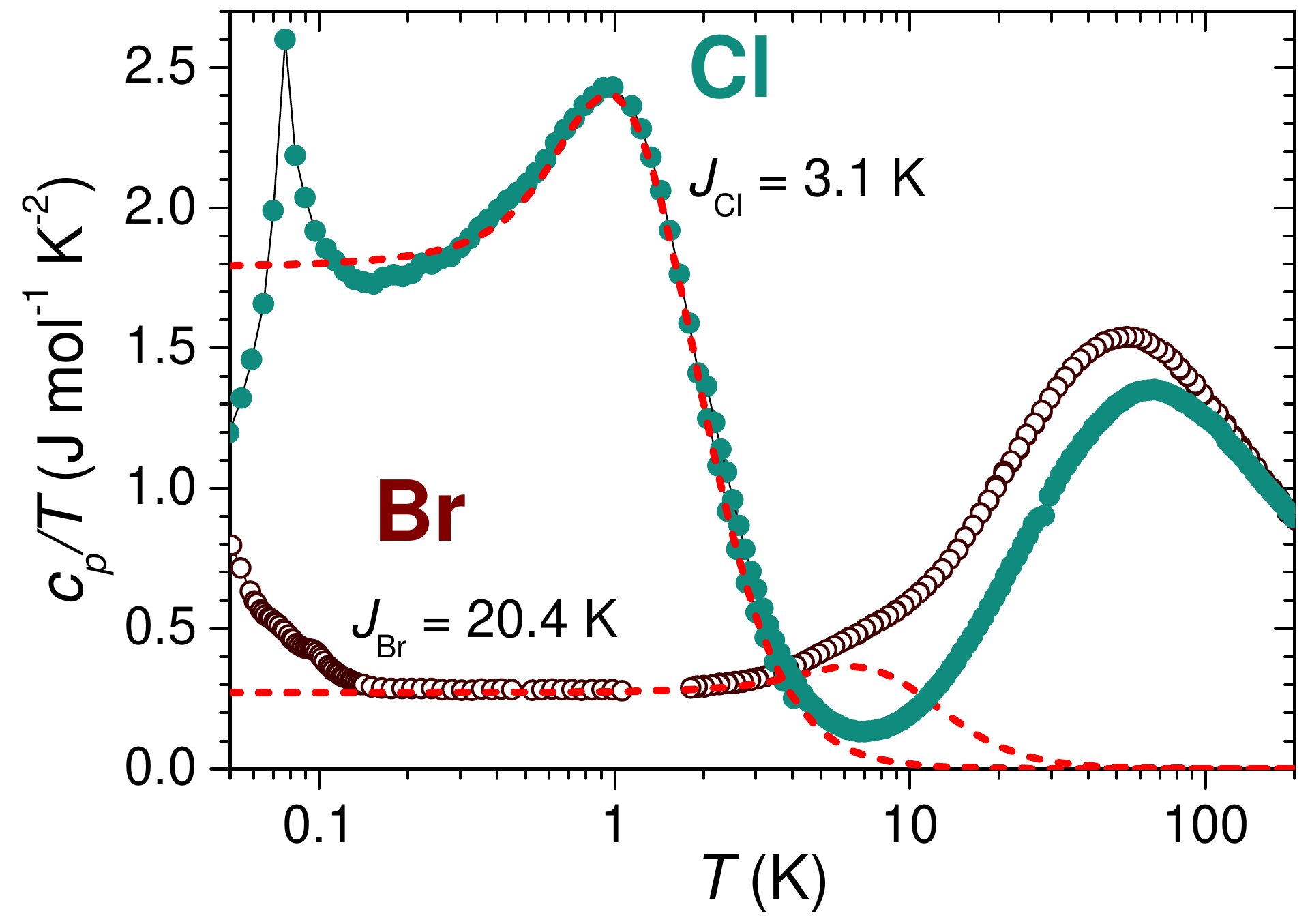}
\caption{Specific heat data in zero field are displayed for the Cl- (filled circles) and Br-compound (hollow circles). Dashed lines correspond to the spin chain specific heat as described in the text.\label{fig::Mandcp-cp}}
\end{figure}
The specific heat data shown in \figref{fig::Mandcp-cp} exhibit three distinct features.
At high temperatures, $c_p/T$ shows the typical phonon peak, where the higher maximum in temperature for the chlorine compound is representative for a higher Debye temperature.
Below 150 mK the onset of magnetic ordering is observed in both compounds, which is described in detail in \secref{sec::PD}. The broad spin chain like feature at intermediate temperatures is described well applying the empirical fitting function to the specific heat of spin chains given by Johnston \etal{}.\cite{Johnston2000} The maximum of the magnetic specific heat in the chlorine compound observed at about 0.95~K is well described with $J_\text{Cl}=3.1(1)\K$. In the bromine compound the peak in the magnetic specific heat is not well separated from the phonon contribution. Note however, that the linear regime in $c_p/T$ between 0.2 and $2\K$ -- as expected for Tomonaga-Luttinger spin liquid like behavior of spin chains -- is coherently well described with $J_\text{Br}=20.4\K$ as determined from susceptibility.
The specific heat of the Cl-compound exhibits an unexpected suppression compared to the spin chain model below 0.3~K, well above the ordering temperature.
\begin{figure}[htb!]
\unitlength1cm
\includegraphics[width=.48\textwidth]{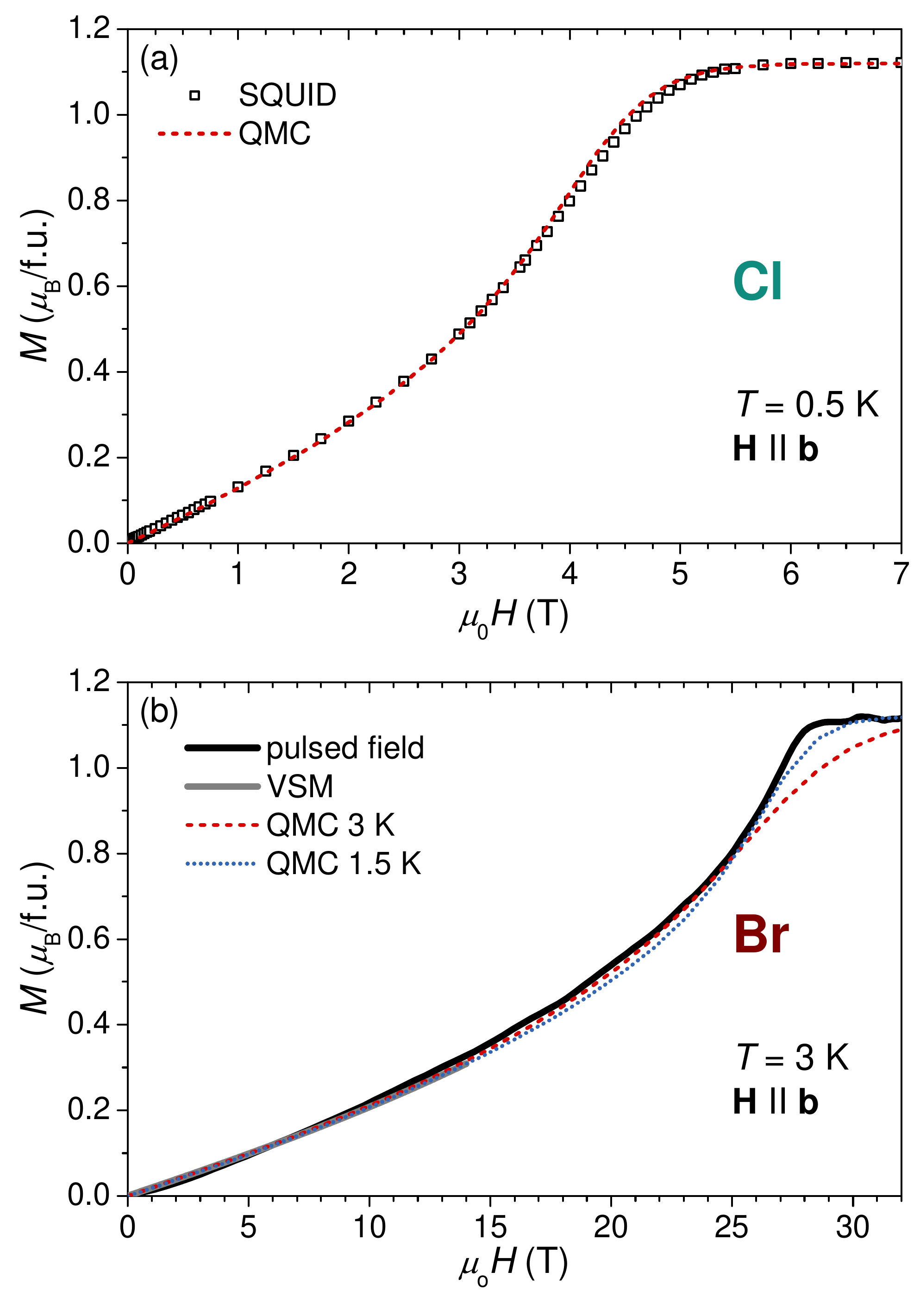} 
\caption{Magnetization data up to the saturation magnetization are shown for \KCSCl{} ((a), SQUID data) and \KCSBr{} ((b), pulsed magnetic field). Dashed lines represent QMC data for spin chains at temperatures as used in the experiments with $g=2.24$, $J_\text{Cl}=3.2\K$ and $J_\text{Br}=20.4\K$. For \KCSBr{}, QMC data are also displayed for $T=1.5\K$ (dotted line).\label{fig::Msat}}
\end{figure}

The energy scale of both compounds is conveniently low to study the magnetization up to the saturation field.
Respective data are presented in \figref{fig::Msat} and compared to quantum Monte-Carlo (QMC) simulations, computed using the ALPS 2.1 package.\cite{Bauer2011}
Magnetic field is applied parallel to the $b$-axis, i.e. parallel to the DM vector and with maximal $g$-factor. QMC data were generated at a $J/T$ ratio corresponding to the temperature of the experiment and using $J$ from the above specific heat data analysis.
The magnetization data of \KCSCl{} are almost perfectly reproduced by the chain-only QMC data.
For \KCSBr{} in contrast, it is found that the susceptibility just below saturation is enhanced compared to the QMC model.
This feature may be attributed to quasi-adiabatic cooling of the sample in the pulsed field, rising to 36 T in 5 ms.
With the sample at $3 \K$ in zero field, its temperature may be significantly reduced with increasing field.
Quasi-adiabatic cooling however can not account for the apparent enhancement of magnetization at intermediate fields, as outlined by comparison with QMC data for $T=1.5\K$.

\subsection{Neutron spectroscopy}

While the thermodynamic data provide evidence that both materials represent spin chain compounds, the predominant exchange path has been detected by inelastic neutron scattering.
As illustrated in \figref{fig::structure}, a first look at the structure suggests strong exchange both in the $a$- and the $b$-direction, the latter potentially frustrated.
The weak interactions and the low saturation field $\mu_0H_{\mr{sat}}<5$~T of \KCSCl{} make the compound well suitable for the determination of the exchange parameters from inelastic neutron scattering data.
In the fully saturated state the excitation spectrum can be analyzed within classical spin wave theory.\cite{Coldea2002}
External magnetic field of $\mu_0H=12\T\gg \mu_0H_{sat}$ was applied in order to align all spins and to guarantee a significant spin gap even at the relatively high sample temperature $T=1.6\K$.
In order to attain information about exchange interactions along any crystallographic direction, the sample has been mounted both in the $(h,k,0)$ scattering plane for measurement of the dispersion along $a^\ast$ and $b^\ast$ and in the $(h,0,l)$ scattering plane for an additional study of the dispersion along $c^\ast$.
Magnetic field was applied perpendicular to the scattering plane.
By energy scans of the incoherent line, the energy resolution of the experiment was determined to be $0.27\meV$ (FWHM).

\begin{figure}[!htb]
\unitlength1cm
\includegraphics[width=.47\textwidth]{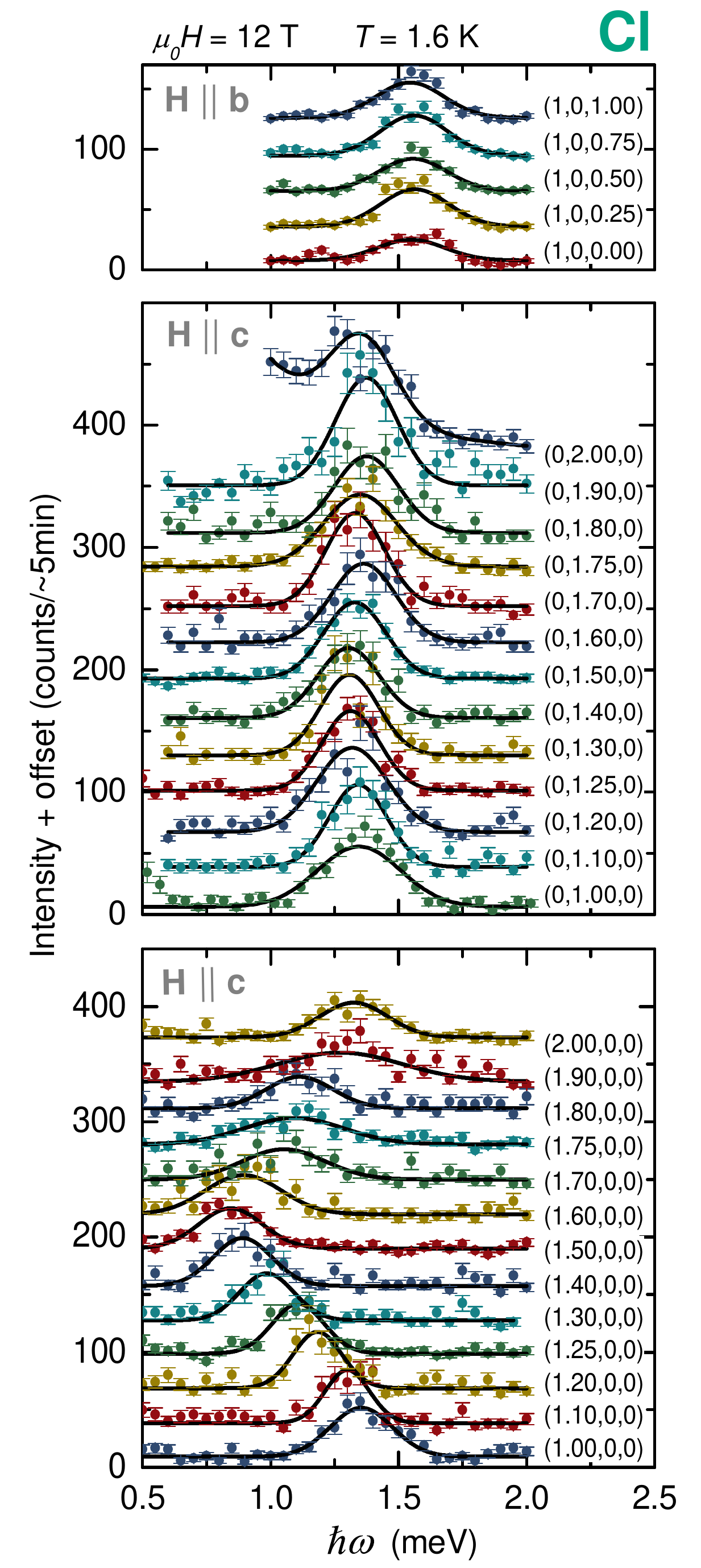} 
\caption{Raw data of the magnon spectrum of \KCSCl{} in the field saturated state at 12~T are displayed for momentum transfers $\textbf{q}=(1,0,l)$, $\textbf{q}=(0,k,0)$ and $\textbf{q}=(h,0,0)$. Solid black lines are Gaussian fits. The nuclear Bragg peak at $(0,2,0)$ is taken into account by an additional exponential term in the fit. The direction of the applied magnetic field is denoted in the figure. In $k$- and $l$-directions no distinct dispersion is detected, but a bandwidth of about $0.5\meV$ was observed along $h$ (see \figref{fig::INS}). \label{fig::INS}}
\end{figure}

Energy transfer scans were performed at constant momentum transfer.
The data show a pronounced dispersion along the $a^\ast$ direction with a bandwidth of about $0.5\meV$.
However, no distinct dispersion in perpendicular directions can be observed.
Detailed inelastic scattering data together with Gaussian fits of the magnetic excitation are shown in \figref{fig::INS}.

\begin{figure}[!htb]
\unitlength1cm
\includegraphics[width=.48\textwidth]{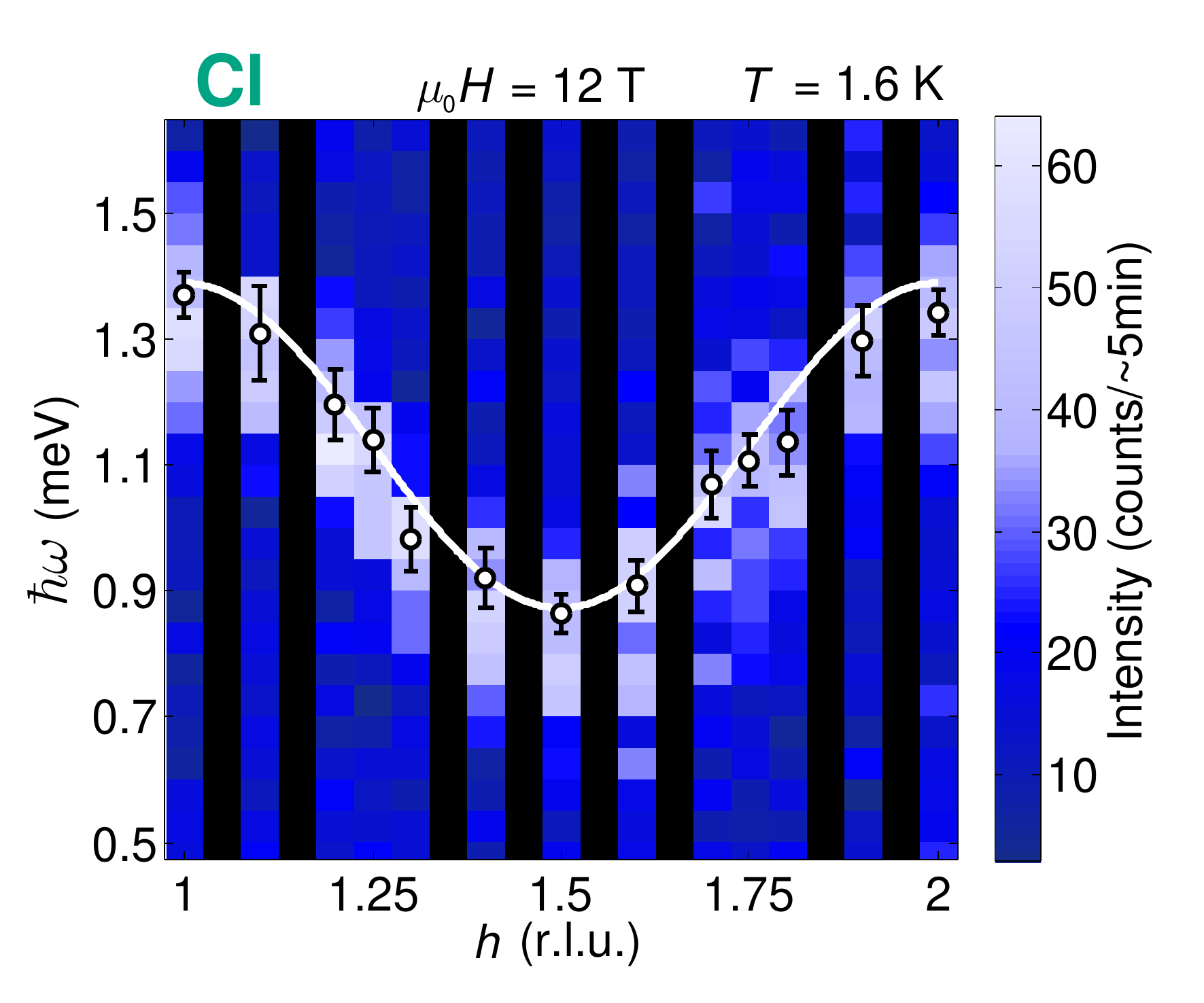} 
\caption{Neutron scattering intensity map of the magnon dispersion along the $a^\ast$-axis ($\textbf{q}=(h,0,0)$) for \KCSCl{} in the fully saturated state ($\mu_0H = $~12~T, $\mb{H}||\mb{c}$). Raw data are shown in \figref{fig::INS}. Dots represent the fitted maximum position of the individual energy scans with respective error bars. The fitted magnon dispersion according to equation (\ref{eqn::dispersion}) is plotted as a white line.\label{fig::INSdisperion}}
\end{figure}

The Heisenberg and Zeeman Hamiltonian of the system reads
\begin{equation}
\mathcal{H}_0=\frac{1}{2}\sum_{i,r} J_r \mb{S}_i\mb{S}_{i+r}-g\mu_0\muB H \sum_i S^z_i\mr{,}
\end{equation}
with magnetic field $H$ and the sum running over all magnetic sites $i$ and exchange paths $r$.
The observed magnon spectrum is found to be described well with a minimal number of exchange constants. These are $J_a$, the predominant exchange along the $a$-axis, the nearest neighbor exchange $J_{b,\mr{nn}}$ along the $b$-axis and an effective exchange $J_{c,\mr{eff}}$, neglecting the difference between $J_c$ and $J_{c'}$.
For detail of the naming convention of the exchange interactions, see \figref{fig::structure}.
The respective spin wave dispersion reads
\begin{eqnarray}
\hbar \omega = g \muB H &-& J_{\mr{Cl},a} (1 - \cos 2 \pi h) - J_{\mr{Cl},b,\mr{nn}} (1 - \cos \pi k) \nonumber\\
                        &-& 2 J_{\mr{Cl},c,\mr{eff}} (2 - \cos \pi h \cos \pi l).\label{eqn::dispersion}
\end{eqnarray}
From a coherent spin wave fit to all data (\figref{fig::INSdisperion}), exchange interactions of $J_{\mr{Cl},a}=2.9(3)\K$, $J_{\mr{Cl},b,\mr{nn}}=0.2(2)\K$ and $J_{\mr{Cl},c,\mr{eff}}=0.0(2)\meV$ have been obtained.
Introducing further exchange constants $J_{\mr{Cl},ab}$, $J_{\mr{Cl},b,\mr{nnn}}$ or distinct exchange paths $J_{\mr{Cl},c}$ and $J_{\mr{Cl},c'}$ yields vanishing exchange for those bonds.
Therefore one can conclude from the observed excitation spectrum that \KCSCl{} exhibits well isolated spin chains along the $a$-axis, despite the fact that nearest neighbor Cu$^{2+}$ ions along the $a$-axis are spatially even more distant than next-nearest neighbors along the $b$-axis.
Indeed, exchange along $b$ involves apical Cu$^{2+}$ orbitals and should hence be strongly suppressed.
Furthermore, no evidence for geometric frustration is found as indicated by the vanishing magnitude of both $J_{\mr{Cl},b,\mr{nnn}}$ and $J_{\mr{Cl},c,\mr{eff}}$.

\begin{figure}[!htb]
\unitlength1cm
\includegraphics[width=.48\textwidth]{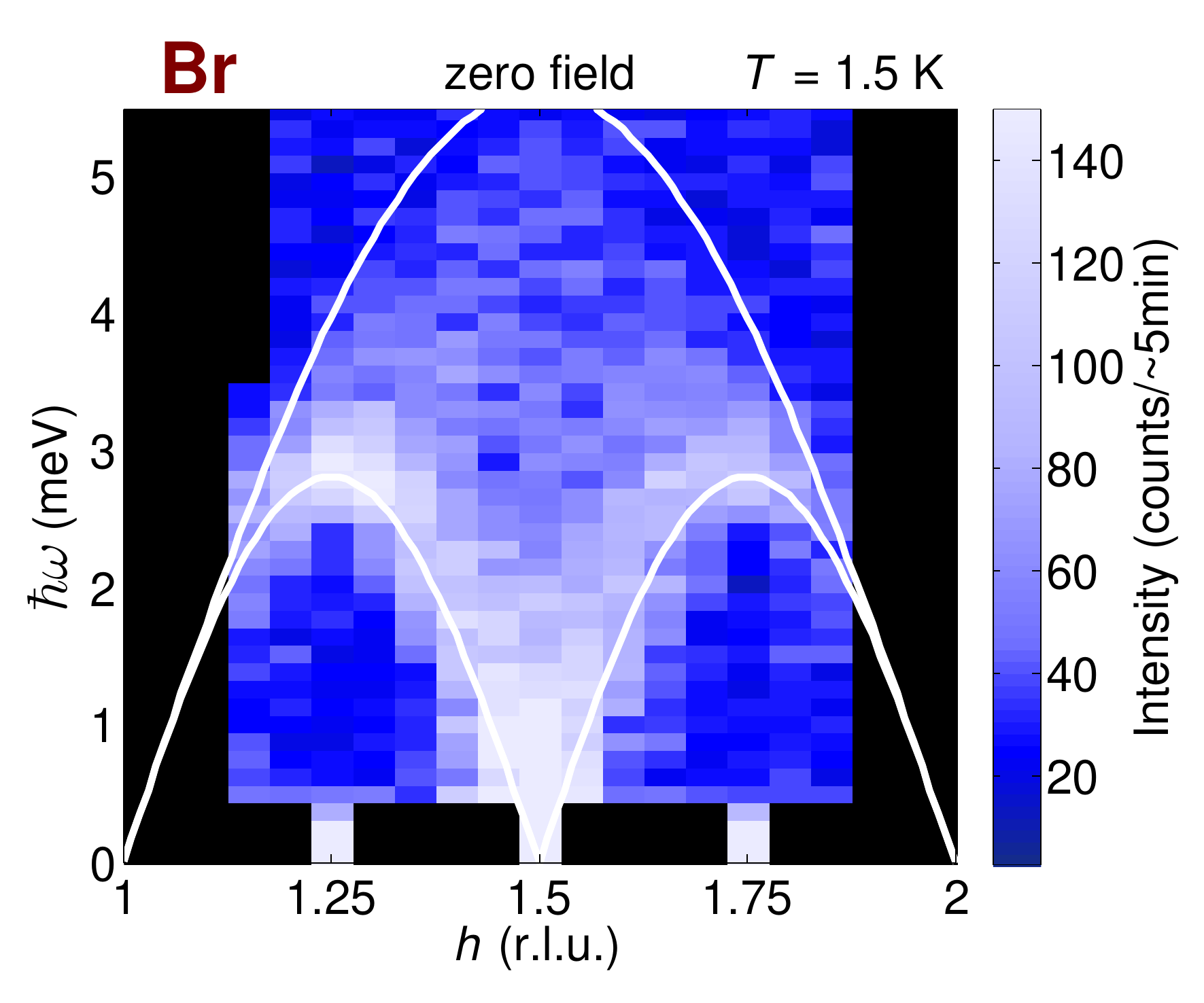}
\caption{Neutron scattering intensity map of the two-spinon continuum along the $a^\ast$-axis ($\textbf{q}=(h,0,0)$) for \KCSBr{} in zero magnetic field. Raw data are shown in \figref{fig::INSBr}. The white line indicates the lower and upper boundary of the continuum following equations (\ref{eqn::epsilon1}) and (\ref{eqn::epsilon2}) with $J_{\mr{Br},a} = 20.7\K$. \label{fig::INSspecBr}}
\end{figure}

\begin{figure}[!htb]
\unitlength1cm
\includegraphics[width=.44\textwidth]{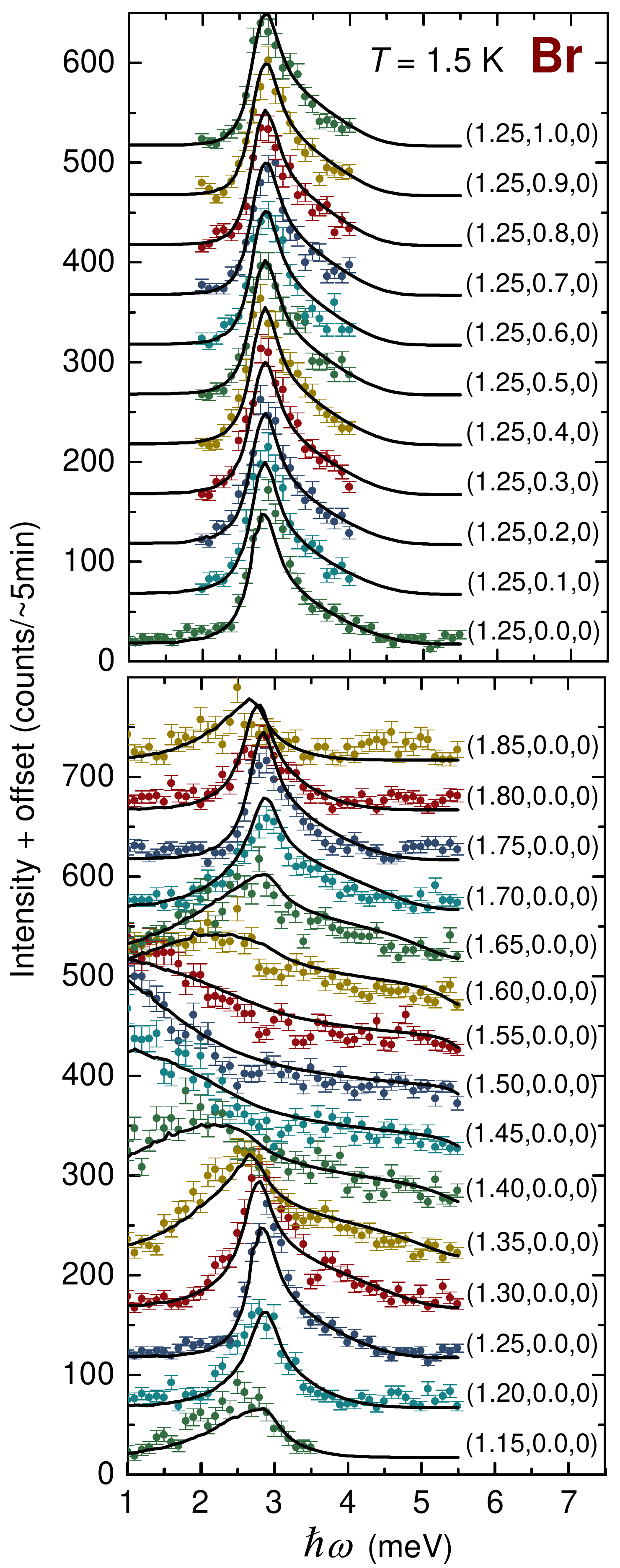}
\caption{Raw data of the spectrum of \KCSBr{} are displayed for momentum transfers $\textbf{q}=(h,0,0)$ and $\textbf{q}=(1.25,k,0)$. Solid black lines are fits of the resolution convoluted model as described in the text. The two-spinon continuum characteristic of $S=1/2$ chains can be observed along $h$ (see \figref{fig::INSspecBr}) whereas no change is visible along $k$. \label{fig::INSBr}}
\end{figure}

Sharing the same crystal structure as the chlorine compound, the bromide is expected to exhibit predominant exchange along the $a$-axis, too.
Inelastic neutron scattering data of \KCSBr{} in zero magnetic field at 1.5~K in the $(h,k,0)$ scattering plane reveal a two-spinon continuum characteristic of $S=1/2$ chains along $a^\ast$ (\figref{fig::INSspecBr}).
The spectrum of $S=1/2$ chains is described well by the model provided by M\"uller \textit{et al.}:\cite{Mueller1981}
\begin{eqnarray}
S\left(h,\omega\right)	&=&	\frac{A}{\sqrt{\left(\hbar\omega\right)^2 - \epsilon_1^2\left(h\right)}} \nonumber\\
												& &	 \times\Theta\left(\hbar\omega-\epsilon_1\left(h\right)\right)\Theta\left(\epsilon_2\left(h\right)-\hbar\omega\right), \label{eqn::spectrum}
\end{eqnarray}
where $\Theta\left(\cdot\right)$ is the step function and $\epsilon_1$ ($\epsilon_2$) denotes the lower\cite{Cloizeaux1962} (upper) boundary of the continuum given by
\begin{equation}
\epsilon_1\left(h\right) = \frac{\pi}{2}J_{\mr{Br},a}\left|\sin{2\pi h}\right|,\label{eqn::epsilon1}
\end{equation}
\begin{equation}
\epsilon_2\left(h\right) = \pi J_{\mr{Br},a}\left|\sin{\pi h}\right|.\label{eqn::epsilon2}
\end{equation}
The coupling constant $J_{\mr{Br},a} = 20.7(2)\K$ was obtained by fitting the correlation function (\ref{eqn::spectrum}) convolved with the instrumental resolution using the ResLib package\cite{Reslib2009} as shown in \figref{fig::INSBr}.
For momentum transfer along $b^\ast$ no distinct changes are discernible (\figref{fig::INSBr}).
Concluding from the bandwidth of the lower boundary of the continuum and the energy resolution, the exchange constants along the $b$-axis are $J_{\mr{Br},b,\mr{nn}} = 0.0(5)\K$ and $J_{\mr{Br},b,\mr{nnn}} = 0.0(5)\K$.

\subsection{Deviations from the Heisenberg Hamiltonian}\label{sec::DM}

\begin{figure}[!htb]
\unitlength1cm
\includegraphics[width=.48\textwidth]{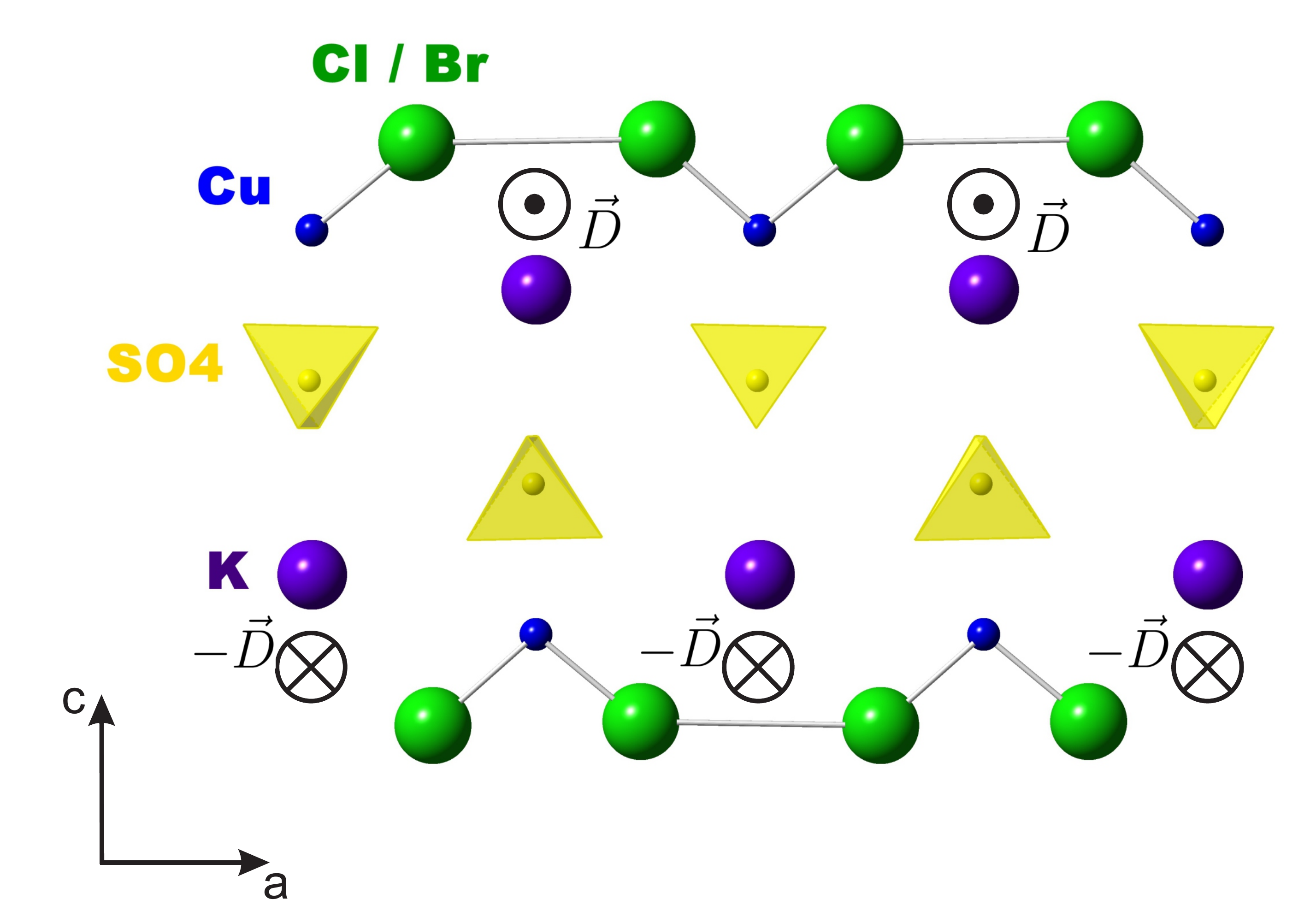} 
\caption{The antisymmetric contribution $\textbf{D}$ to $J_a$ points along the $b$-axis, is uniform along the chain and antiparallel in adjacent chains. \label{fig::DMstructure}}
\end{figure}

From the magnon dispersion a single predominant exchange path $J_a$ along Cu-Ha-Ha-Cu bonds was found.
The exchange interaction is mediated by two halide ions that are located far off the axis connecting the respective Cu$^{2+}$-ions (cf.~\figref{fig::structure}).
Thus, the isotropic exchange should be associated with an antisymmetric component.
A symmetry analysis of the crystal structure shows that a \DM{} contribution $\textbf{D}$ is expected for the chain running along $a$.
The \DM{} vector $\textbf{D}$ has to point exactly along $b$ and is uniform along the chain and antiparallel in neighboring chains as illustrated in \figref{fig::DMstructure}.\cite{Dzyaloshinski1958,Moriya1960}

Correspondingly, the magnetic properties of the materials in question can be discussed in terms of the following anisotropic Hamiltonian:
\begin{eqnarray}
\mathcal{H}	&=& \sum_{i,j,k} J_a \mb{S}_{i,j,k}\mb{S}_{i+1,j,k} \nonumber \\
						&+& \sum_{i,j,k} (-1)^{j+k}D \left( S^z_{i,j,k} S^x_{i+1,j,k} - S^x_{i,j,k} S^z_{i+1,j,k} \right),
\end{eqnarray}
where $i, j, k$ run over the sites along the $a, b, c$ axis.

\subsection{ESR spectroscopy}

Electron spin resonance (ESR) was recently predicted\cite{Gangadharaiah2008} and experimentally demonstrated\cite{Povarov2011,Fayzullin2013} to be a sensitive probe for DM interaction in spin chain systems. In the presence of DM interactions the excitation spectrum of the spin chain is shifted along the spinon propagation direction and develops a corresponding gap at zero momentum transfer.
In magnetic field and at $T=0$~K, ESR resonances are to be observed at frequencies $h\nu=|g\mu_0\muB \textbf{H}\pm \frac{\pi}{2}\textbf{D}|$.\cite{Povarov2011} For magnetic field applied along the DM vector $\textbf{D}$, the ESR line is split by $\mu_0\Delta H=\pi D/\left(g\muB\right)$, but only a single shifted line is expected for magnetic field perpendicular to $\textbf{D}$.

\begin{figure}[!htb]
\unitlength1cm
\includegraphics[width=.48\textwidth]{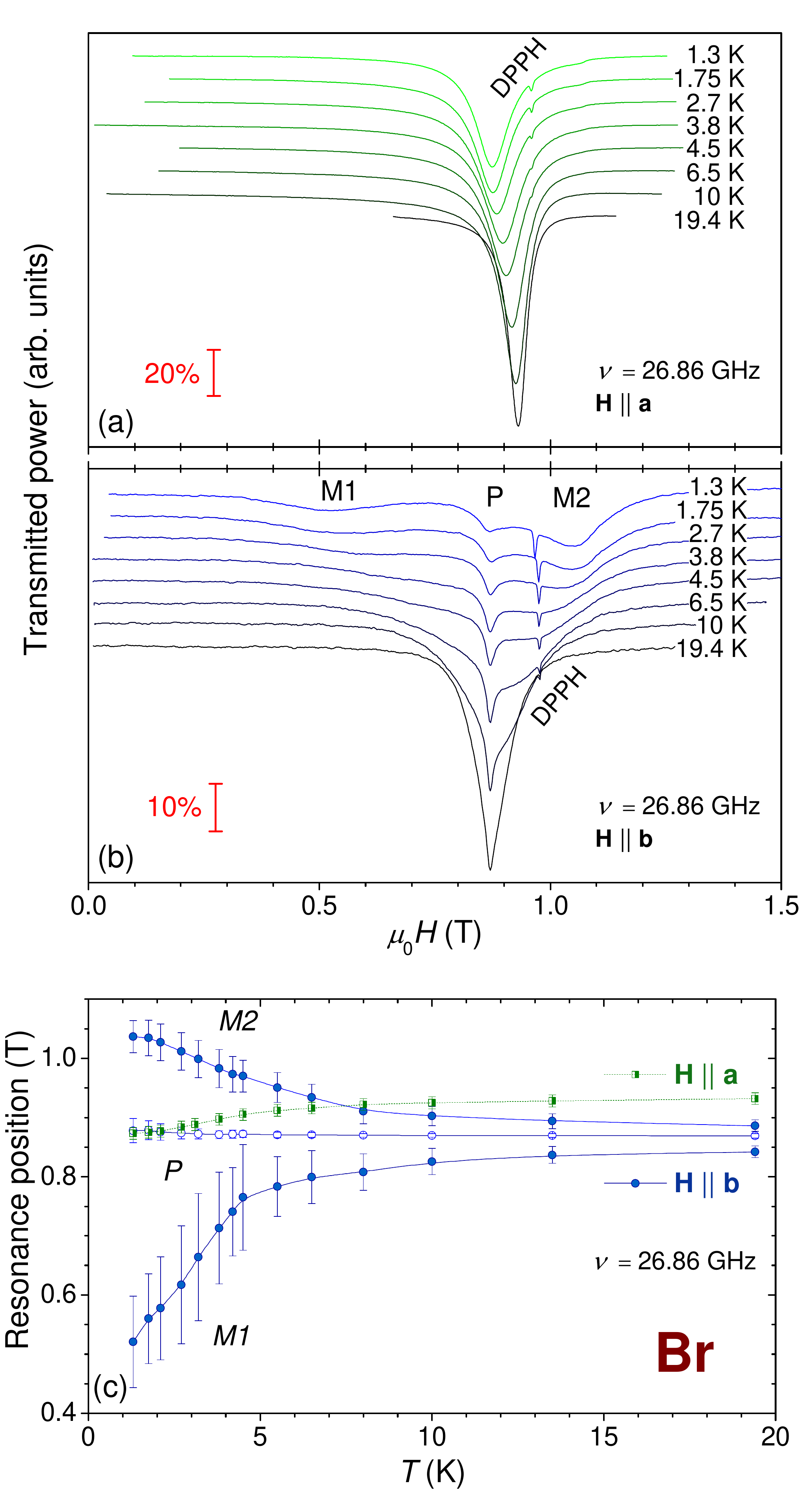}
\caption{ESR spectra of \KCSBr{} for magnetic field perpendicular (a) and parallel (b) to the DM vector ($b$-axis of the crystal) for various temperatures. In the latter case, the line-shape up to $19.4\K$ must be described as a sum of three Lorentzians. The temperature dependence of the resonance line positions is plotted in (c). Error bars are proportional to the fitted line widths. DPPH was used for calibration giving rise to the additional signal indicated in the figure. \label{fig::ESRraw}}
\end{figure}

Temperature dependent ESR data have been measured on \KCSBr{} -- selected for its larger intra-chain exchange -- with magnetic field applied in the $ab$-plane.
Data were recorded at temperatures between 1.3~K and 19.4~K, a subset of which are shown in \figref{fig::ESRraw}.
For field applied along the $a$-axis, i.e. perpendicular to $\textbf{D}$, a single absorption line is observed.
Its resonance magnetic field decreases monotonously with temperature.
In contrast, three lines are observed for field along the $b$-axis (parallel to $\textbf{D}$).
The central line, denoted as $P$, does not exhibit any notable shift with temperature.
In contrast, the lines labeled $M_1$ and $M_2$ become well separated at low temperatures and can still be distinguished up to highest temperatures in the experiment $\kB T\approx J_a$.
All data are well described with a Lorentzian line shape.
The temperature dependence of the fitted resonance fields as well as the respective line-widths are summarized in \figref{fig::ESRraw}.

In summary, uniaxial ESR line-splitting is observed for field along the $b$-axis as well as a line-shift for field applied in a perpendicular direction.
Following the argument and analysis of Povarov \etal{},\cite{Povarov2011} the ESR data are strong indications of the presence of antisymmetric exchange with uniform DM vector pointing along the $b$-axis. The observed line splitting $\mu_0\Delta H = 0.51(9)\T$ at $1.3\K$ for field parallel to the $b$-axis allows to estimate $D=0.24(5)\K$ with $g_b=2.24$, as obtained from the (central) line at $19.4\K$. At the same time, one obtains $D= 0.31(5)\K$ from the resonance field $\mu_0H=0.87(2)\T$ measured at $1.3\K$ for field parallel to the $a$-axis ($g_a=2.04$). Both estimates of $D$ are significantly larger than the ordering temperature $\TN \approx 100\mK$ of \KCSBr{}.

However, this model\cite{Gangadharaiah2008} for the ESR data does not account for the observed temperature-independent $P$-line. As a corresponding temperature-independent line for $\mb{H}||\mb{a}$ is absent, the $P$-line cannot be attributed to potential paramagnetic impurities. Studies with polarized neutrons in the paramagnetic regime may shed further light on the precise magnitude of the DM interaction.\cite{Aristov2000,Derzhko2006}

\subsection{Magnetic phase diagram}\label{sec::PD}
\begin{figure}[!htb]
\unitlength1cm
\includegraphics[width=.48\textwidth]{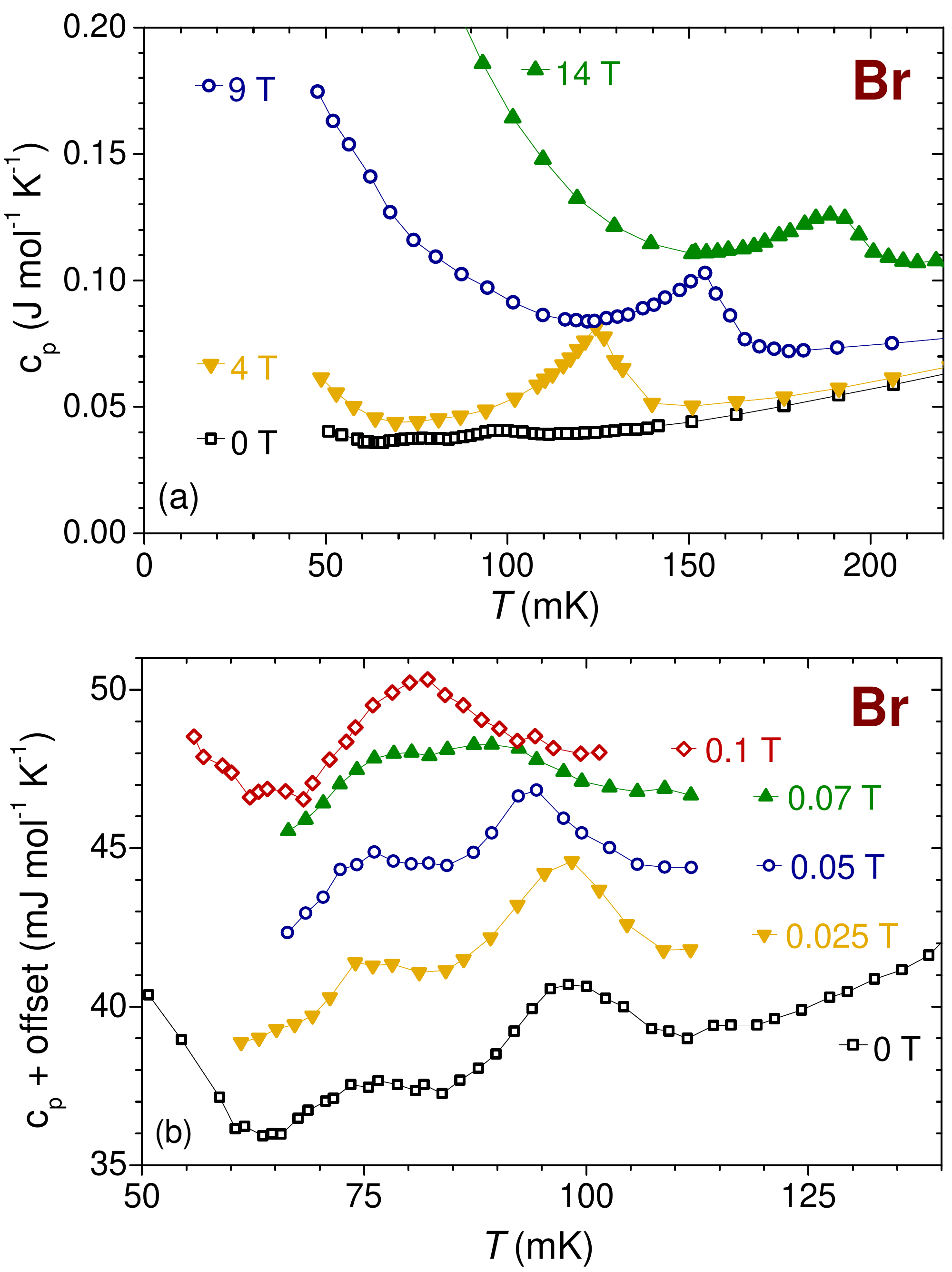}
\caption{Low temperature specific heat data of \KCSBr{} in magnetic field $\mb{H}||\mb{b}$. Data for magnetic field below $0.1\T$ is shown in (b) applying a constant offset of 2.5~mJ~mol$^{-1}$~K$^{-1}$ between data at different magnetic fields. Two separate maxima are observed in zero magnetic field which merge below $0.1\T$. \label{fig::cpHBr}}
\end{figure}

\begin{figure}[!hbt]
\unitlength1cm
\includegraphics[width=.4\textwidth]{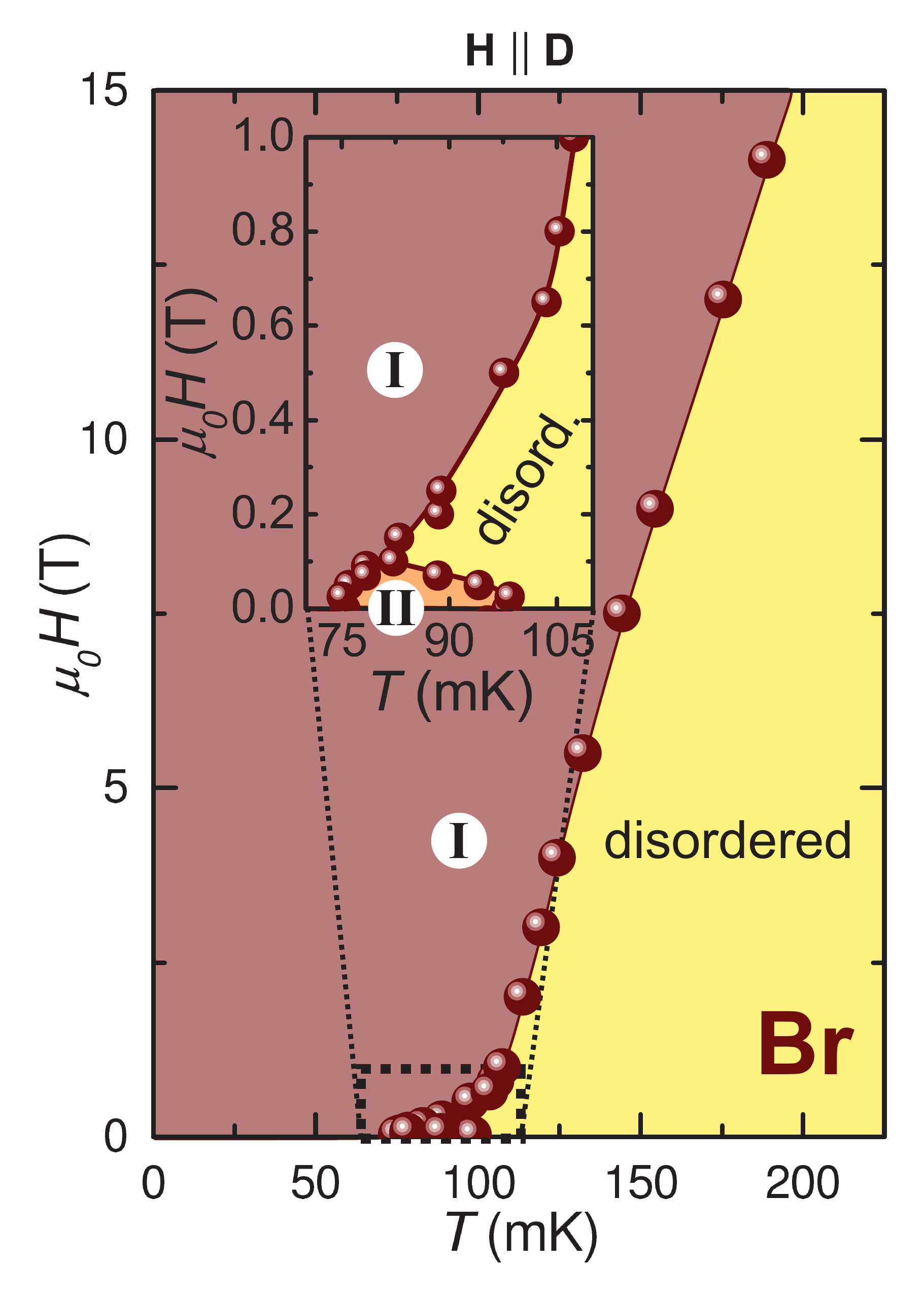}
\caption{Magnetic phase diagram of \KCSBr{} for magnetic field applied along the $b$-axis, parallel to the DM vector. Highlighted regions are guides to the eye, denoting different phases. The inset shows the low field phase (phase II) observed in \KCSBr{} only. Phases I and II are interpreted as two distinct magnetically ordered phases. \label{fig::PDBr}}
\end{figure}

The magnetic phase diagrams of both compounds are studied by specific heat measurements at temperatures down to $50\mK$ and magnetic fields up to $14\T$ applied along the $b$-axis, i.e. parallel to the DM vector.
From the presence of a sizable DM interaction, as indicated by ESR, a complex magnetic phase diagram of \KCSBr{} is expected in particular for field non-parallel to $\textbf{D}$.\cite{Muhlbauer2012}
However, in this study, the focus is on fields parallel to $\textbf{D}$, where specific heat data already reveal a remarkable behavior of \KCSBr{}.
Data are presented in \figref{fig::cpHBr}.
In zero magnetic field, no anomaly indicating magnetic ordering is observed down to about $150\mK$.
At lower temperatures two weak yet distinct maxima at $75\mK$ and $100\mK$ are observed, well different from a lambda-like ordering feature.
Below $65\mK$ the specific heat grows significantly, indicative for nuclear specific heat.
The latter anomaly increases substantially with applied field, as to be expected for a nuclear Schottky anomaly.
Its presence already in zero magnetic field suggests that the two maxima might indeed be interpreted as the signature of static magnetic order.
Upon increasing magnetic field up to $0.1\T$ the two broad features continuously merge and at higher fields a single phase transition is clearly detected by a typical lambda-like maximum in the specific heat.
For all curves, the observed maxima in $c_p(T)$ are interpreted as the critical temperature.
The data are summarized in the magnetic phase diagram presented in \figref{fig::PDBr}.
Between $0.1\T$ and $14\T$ phase I is stabilized with respect to temperature, as is common for the ordered phase of low-dimensional magnets in intermediate magnetic fields.\cite{DeGroot1986,Giamarchi1999}
The two broad low field features appear to enclose a distinguished finite-temperature magnetically ordered phase (phase II) which is rapidly suppressed by magnetic field.

\begin{figure}[!htb]
\unitlength1cm
\includegraphics[width=.48\textwidth]{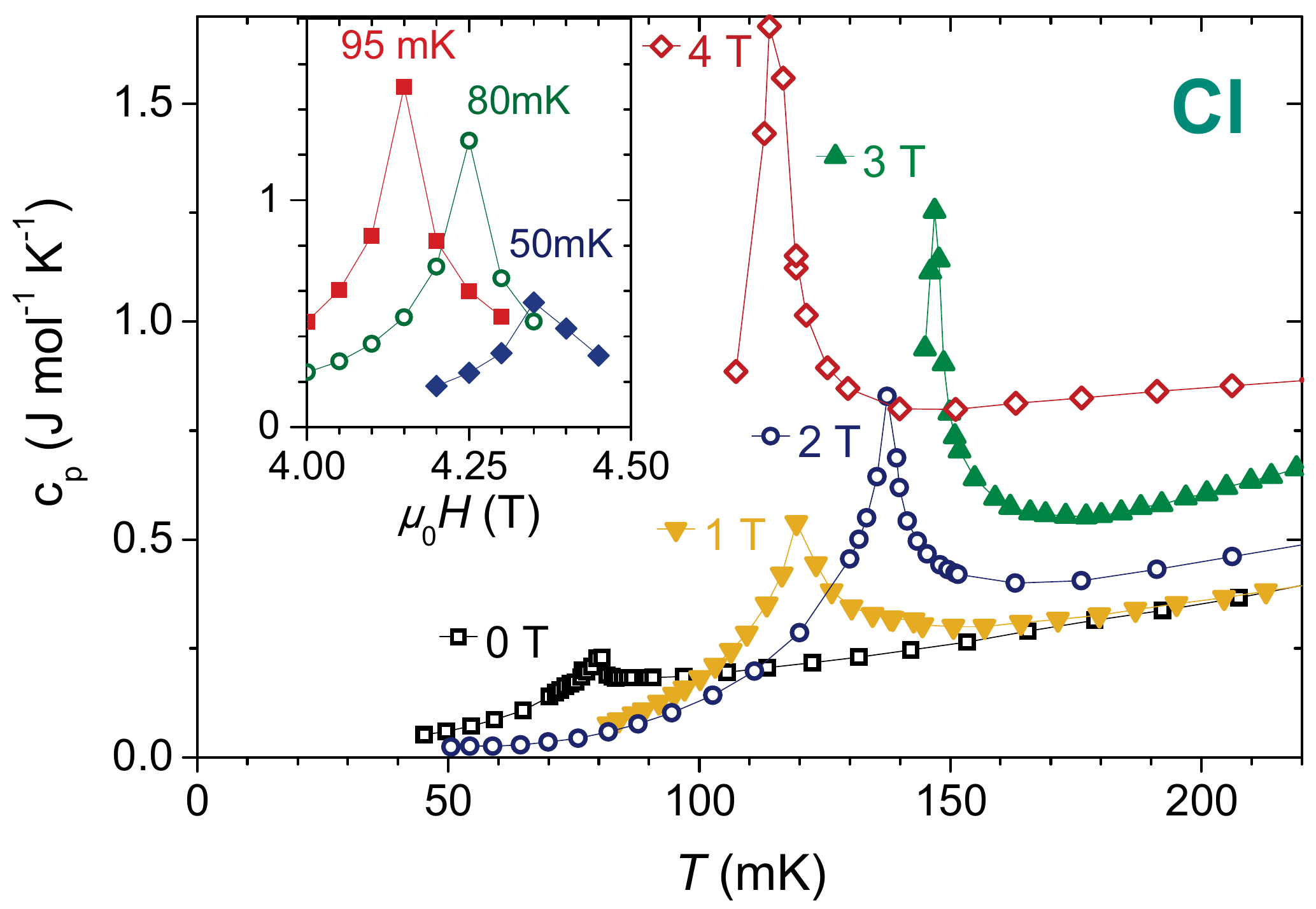}
\caption{Low temperature specific heat data of \KCSCl{} in magnetic field $\mb{H}||\mb{b}$. The inset displays the specific heat of \KCSCl{} measured at constant temperature versus field. \label{fig::cpHCl}}
\end{figure}

\begin{figure}[!hbt]
\unitlength1cm
\includegraphics[width=.4\textwidth]{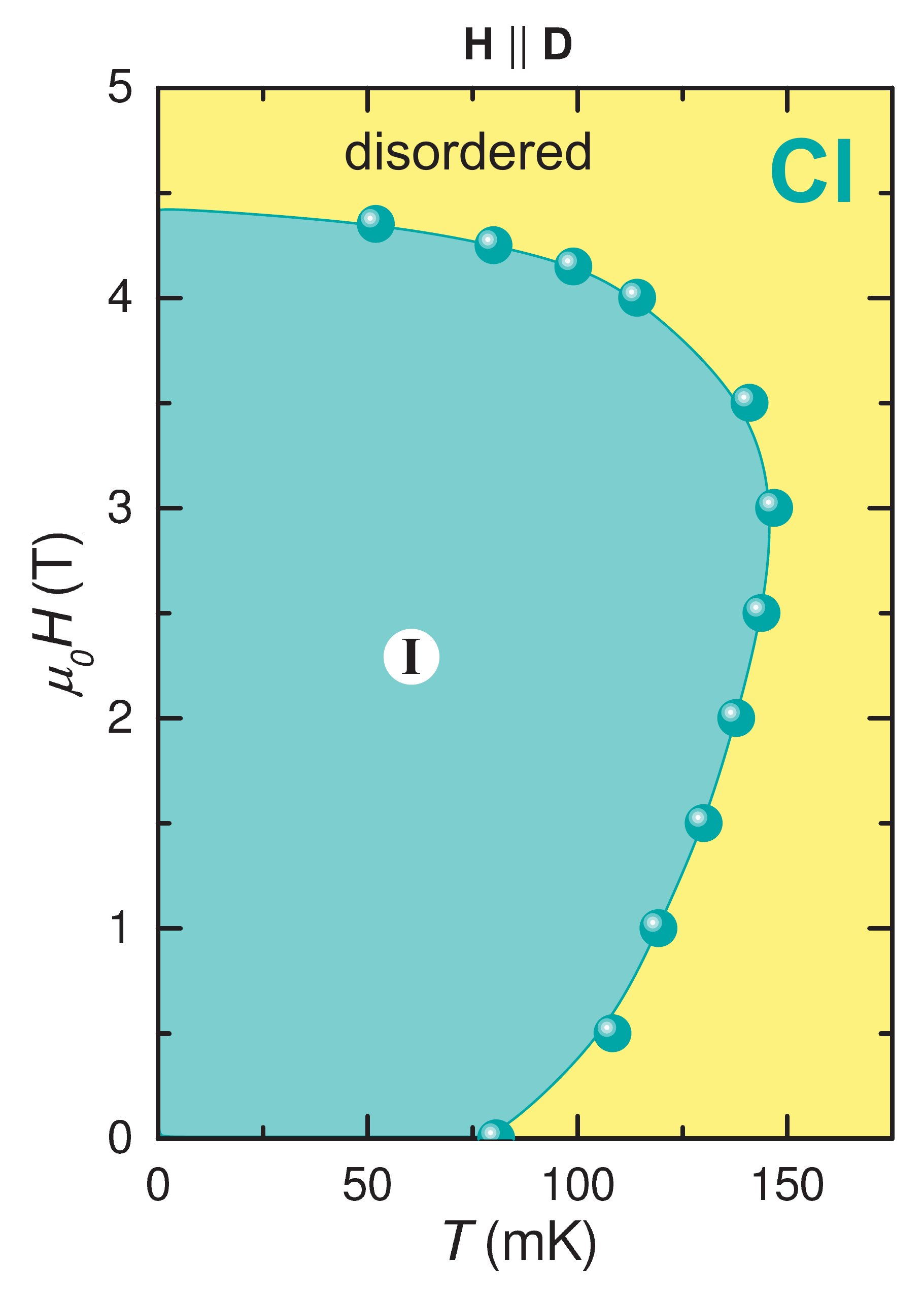}
\caption{Magnetic phase diagram of \KCSCl{} for magnetic field applied along the $b$-axis, parallel to the DM vector. Highlighted regions are guides to the eye, denoting different phases. Phase I is interpreted as a magnetically ordered phase. \label{fig::PDCl}}
\end{figure}

In contrast to the unconventional behavior of \KCSBr{}, \KCSCl{} displays a well defined lambda-like specific heat anomaly even in zero field at $77\mK$.
The phase diagram was studied up to saturation at $4.5 \T$ and respective data are displayed in \figref{fig::cpHCl}.
At lowest temperatures and highest fields, the phase boundary was measured at 'fixed' temperature, changing magnetic field.
The phase transition is assigned to the maximum of the observed peak.
The phase diagram is displayed in \figref{fig::PDCl}.
The observed transition field of $4.35\T$ at $50\mK$ agrees well with the magnetization data of \figref{fig::Msat}.
Note that the latter measurement is performed at a finite temperature of $500\mK$, in the paramagnetic state. Due to thermal fluctuations, the resulting magnetization curve levels off slightly above the projected saturation field for $T\rightarrow 0$.
In contrast to \KCSBr{}, no intermediate phase is observed at low magnetic field in \KCSCl{}.

\section{Discussion}

To discuss the magnetic properties of the two compounds, the hierarchy of the relevant magnetic interactions shall be reviewed first.
To the first approximation, \KCSCl{} and \KCSBr{} are Heisenberg $S=1/2$ chain materials, albeit with considerably different exchange constants $J_\mr{Cl} = 3.1\K$ and $J_\mr{Br} = 20.5\K$, respectively (see table~\ref{tab:const}).
Inter-chain coupling is weak, but can not be neglected, as it is responsible for 3-dimensional ordering.
While the corresponding exchange constants are too small to be measured in neutron experiments, they can be estimated from the ordering temperatures and the intra-chain coupling strengths using chain mean field theory.\cite{Schulz1996,Yasuda2005}
Inter-chain coupling constants of $J'_\mr{Cl}=31\mK$ and $J'_\mr{Br}=34\mK$ were obtained for the chloride and bromide, respecively.
The immediate conclusion is that the Br-compound is considerably more one-dimensional than the Cl-based one.
As illustrated by the ESR experiments, an additional relevant energy scale is provided by intra-chain uniform DM interactions.
In the Br-compound, they constitute 1.4\% of $J$.
The estimates for $J$, $J'$ and $D$ are summarized in table~\ref{tab:const}.
\begin{table}[!htb]
	\centering
		\begin{tabular}{c c c}
			\hline
			\hline
				&	\KCSCl{} & \KCSBr{} \\
			\hline
			$J$ & 3.1(1) K & 20.5(1) K \\
			$J'$ & 0.031(2) K & 0.034(2) K \\
			$D$ & $\sim 0.04$~K & 0.28(5) K \\
			\hline
			\hline
		\end{tabular}
	\caption{Exchange constants for \KCSCl{} and \KCSBr{}: The intra-chain coupling $J$ is the average of the values obtained by thermodynamic and neutron measurements. For the inter-chain coupling $J'$, the formula from Yasuda \textit{et al.}\cite{Yasuda2005} was followed. The two values for $D_\mr{Br}$ from ESR were averaged. For the chloride, $D$  is crudely estimated assuming the same $D/J$ ratio as in \KCSBr{}.\cite{footnote}}
	\label{tab:const}
\end{table}

\begin{figure}[!htb]
\unitlength1cm
\includegraphics[width=.48\textwidth]{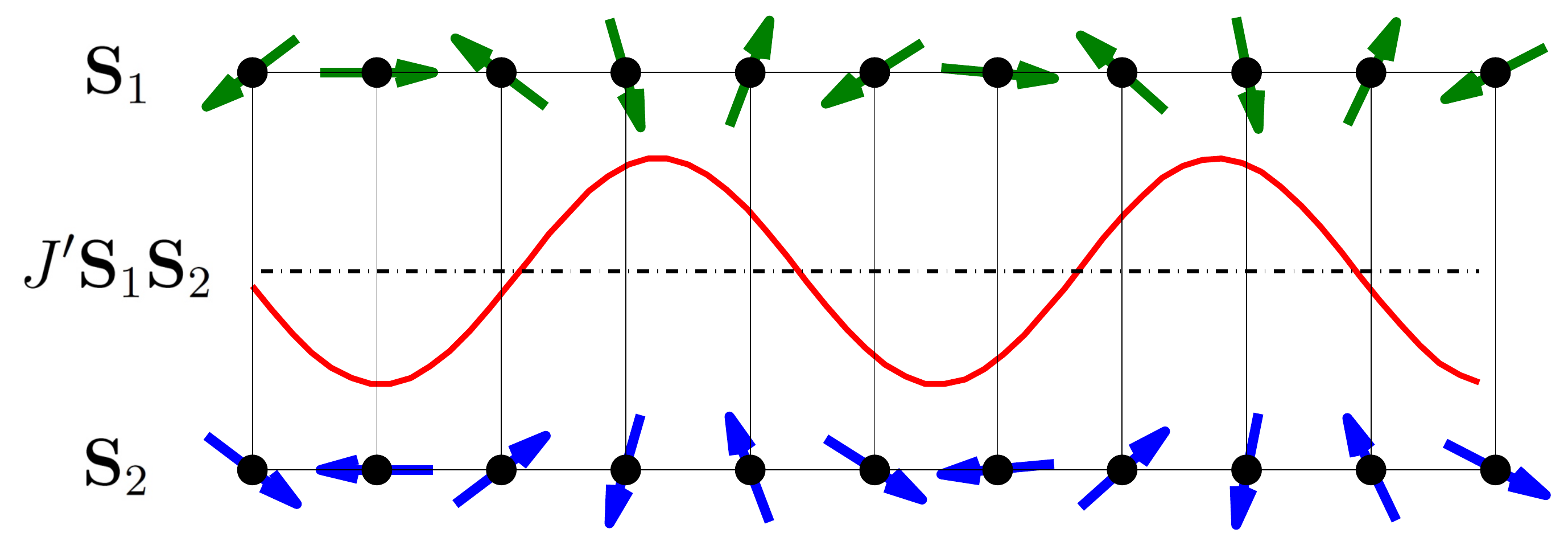}
\caption{Illustration of frustration caused by DM interactions along the chain with inter-chain interactions $J'$ for spin components perpendicular to $\textbf{D}$. As $\textbf{D}$ is antiparallel in nearest-neighbor chains, these exhibit helical correlations with opposing helicity. The effective inter-chain coupling $J'\sum_i \textbf{S}_{1,i} \cdot \textbf{S}_{2,i}$ vanishes for transverse spin components.\label{fig::frustration}}
\end{figure}

For a \emph{single} spin chain, the effect of the DM term would simply be to favor a helimagnetic spin arrangement, with the spiral period determined by $D/J$.
However, for \emph{interacting} spin chains the situation may be more nuanced, as illustrated in \figref{fig::frustration}.
The key issue is that in \KCSCl{} and \KCSBr{} the DM vectors are aligned \emph{antiparallel} to each other in adjacent chains, so that the single-chain spiral correlations tend to propagate in \emph{opposite} directions.
As a result, at the mean field level, inter-chain interactions due to $J'$ are averaged to zero.
This is a peculiar instance of geometric frustration induced not by competing exchange, but by competing DM interactions.
Whether or not this frustration will affect magnetic ordering will depend on how $D$ compares to $J'$.
For small enough $D$, $J'$ is expected to overcome the frustration, and the ground state will simply exhibit commensurate N$\acute{\mr{e}}$el order.
It is likely that this scenario is realized in the chloride.
In contrast, in the bromide, DM interactions are dominant over $J'$.
One can only speculate that this circumstance is at the origin of the additional finite-temperature phase, which may be helimagnetic or perhaps even longitudinally modulated.
The best chance for understanding the phase diagram and the observed finite-temperature phase are future neutron diffraction or local probe experiments.

\section{Conclusions}

In this paper we present the first experimental data on the magnetic properties of the Cu$^{2+}$ salts \KCSCl{} and \KCSBr{}.
Both compounds are isostructural and are well described as spin chain systems as was pointed out by measurements of magnetization and specific heat as well as by inelastic neutron experiments.
Furthermore, the compounds feature peculiar DM interactions which are uniform along the chain and antiparallel in neighboring chains.
This leads to frustration and manifests itself already at accessible low temperatures, for example as extraordinary additional phases.

\section{Acknowledgments}
This work was supported by division II of the Swiss National Fund.
Parts of this work are based on experiments performed at the Swiss spallation neutron source SINQ, Paul Scherrer Institute, Villigen, Switzerland.
Furthermore, we acknowledge the support of the HLD at HZDR, member of the European Magnetic Field Laboratory (EMFL).


\centerline{***}

\bibliography{bibliography}\label{Cbib}

\end{document}